\documentstyle[aps,prl,multicol,floats,epsfig]{revtex}

\begin{document}

\draft

\title{Charge-Induced Fragmentation of Sodium Clusters}

\author{P. Blaise, S. A. Blundell, C. Guet,\cite{claudeaddr} 
and Rajendra R. Zope}

\address{
CEA-Grenoble, D\'{e}partement de Recherche Fondamentale sur
la Mati\`{e}re Condens\'{e}e \\
17, rue des Martyrs, F-38054 Grenoble CEDEX 9, France }

\date{\today}

\maketitle

\begin{abstract}
The fission of highly charged sodium clusters with fissilities $X>1$
is studied by {\em ab initio} molecular dynamics.
Na$_{24}{}^{4+}$ is found to undergo predominantly sequential 
Na$_{3}{}^{+}$ emission
on a time scale of 1~ps, while Na$_{24}{}^{Q+}$ ($5 \leq Q \leq 8$)
undergoes multifragmentation on a time scale $\geq 0.1$~ps, 
with Na$^{+}$ increasingly the dominant fragment as $Q$ increases.  
All singly-charged fragments Na$_{n}{}^{+}$ up to size $n=6$ are 
observed.
The observed fragment spectrum is, within
statistical error, independent of the temperature $T$ of the parent cluster
for $T \leq 1500$~K.  These findings are consistent with and explain
recent trends observed experimentally.
\end{abstract}


\begin{multicols}{2}
    
Starting with the work of Sattler {\em et al.} \cite{Sattler81} on 
van der Waal's clusters, the study of the fragmentation (fission) of 
charged atomic or molecular clusters has proved a valuable experimental 
tool for investigating the intrinsic stability and binding forces of 
these objects.  These studies and others yield important insights 
into the behavior of matter at the small size limit and the size-dependent
evolution toward bulk properties.  Much recent work on fission has 
been devoted to metallic clusters, both experimentally
\cite{Brech91,TP92,Brech94,Brech94LiK,Brech96entrop,Brech98,%
Grenoble95,Guet96,Guet97,Bergen98}
and theoretically 
\cite{Brech94,Guet96,Sven97,Bar91a,GrossIII,Li97,Calvay98}.
Fission of metallic clusters is particularly interesting on 
account of the similarities and differences with the
nuclear fission process \cite{Sven97}.  Over a century ago, 
Lord Rayleigh \cite{LordRayleigh}
studied the problem of the time development of the Coulomb instability
of a surface-charged liquid drop, and conjectured that the excess charge
would be removed by the emission of jets, rather than by fission into
two parts of more or less equal size (as occurs for nuclei, where 
the charge is distributed uniformly through the volume).

According to the Rayleigh criterion \cite{LordRayleigh}, a charged
liquid drop is unstable against Coulomb forces when its fissility
$X > 1$, where $X = E_{\rm Coul}/(2 E_{\rm surf})$ is proportional to the
ratio of the Coulomb to surface energy of the drop.  For a metallic cluster 
M$_{N}{}^{Q+}$, assumed spherical with radius $r_{s} N^{1/3}$  
($r_{s}$ is the Wigner-Seitz radius for the metal), this gives 
$X=(16\pi r_{s}^3 \sigma)^{-1} Q^2/N$, where $\sigma $ is the surface
tension, or $X \approx 2.5 Q^2/N$ for Na$_{N}{}^{Q+}$.  Now, fission experiments
on metallic clusters where the cluster is charged by laser ionization
\cite{Brech91,TP92,Brech94,Brech94LiK,Brech96entrop,Brech98}
have so far produced only clusters with $X < 1$, for which an energy 
barrier exists against fission. Thermally activated fission may
be observed, however, and the dominant charged fragment is found to be
Na$_{3}{}^{+}$, which has a closed electronic shell and is particularly 
stable; Na$^{+}$ has not so far been observed.  In a different type 
of experiment
\cite{Grenoble95,Guet96,Guet97,Bergen98},
sodium clusters are ionized by collision with a beam of
highly charged ions, a technique that is expected to allow study of 
a much wider range of $X$ and cluster temperatures.  Coincidence 
measurements reveal multifragmentation processes in some cases, often 
with Na$^{+}$ as the dominant fragment.

In this Letter, we offer the first systematic dynamical
study of metallic cluster fission in the regime $X >1$ using {\em ab initio} 
molecular dynamics (MD) \cite{abinit}.
We find that for $X$ close to unity, sequential emission of mainly
Na$_{3}{}^{+}$ is the dominant decay channel, while 
for $X \gg 1$, multifragmentation with Na$^{+}$ as the dominant fragment
occurs.  Our simulations provide detailed spatial and temporal information
on the fission process, and explain of some of the trends observed 
experimentally.

On each time step of the cluster dynamics, we compute the 
density of valence electrons (and hence the forces on the ions) 
within the temperature-dependent Kohn-Sham (KS) formalism \cite{Parr}, 
using the local density
approximation (at zero temperature) for the exchange-correlation
functional $E_{xc}[\rho]$.  We use a real-space finite-difference 
method, recently 
developed by us \cite{BlaiseThesis}, incorporating a novel system of 
adaptive simulation cells that surround, adapt to, and follow each 
distinct fragment during a multifragmentation, permitting the simulation 
to continue efficiently up to large fragment separations.  To achieve 
better fragmentation statistics, at the expense of some loss of 
first-principles accuracy, we employ a soft, phenomenological 
pseudopotential \cite{Blaise97}, which permits a relatively large 
real-space grid step $\Delta = 1.35$~a.u.  
We do not expect our choice
of functional $E_{xc}[\rho]$ or pseudopotential to affect significantly
the main results for barrierless fission.
 
To study the fragmentation of a single species Na$_{N}{}^{Q+}$ at an 
``initial temperature'' $T_{\rm in}$, we run $M$ dynamical simulations
arising from an ensemble of $M$ initial conditions constructed 
as follows: (i) We optimize the geometry of the neutral cluster
Na$_{N}$ at 0~K; (ii) we perform an MD run of $\geq 20$~ps for
Na$_{N}$ at 400~K (which is roughly the temperature of the clusters 
Na$_{N}$ output by the cluster source in the collision experiments
\cite{Grenoble95,Guet96,Guet97,Bergen98});
(iii) we take $M$ ionic (nuclear) configurations 
$\{{\bf R}_{I}^{(n)}\}$ with
velocities $\{{\bf V}_{I}^{(n)}\}$ ($n=1$--$M$) at regular intervals 
from this simulation; (iv) for each $\{{\bf R}_{I}^{(n)}\}$, we remove
$Q$ electrons and 
\end{multicols}

\twocolumn

\noindent re-equilibrate the remaining electrons to an
electronic temperature $T_{\rm el} = T_{\rm in}$; and (v) we start the
dynamics with configuration $\{{\bf R}_{I}^{(n)}\}$ and
velocities $\{\lambda {\bf V}_{I}^{(n)}\}$, with $\lambda$ chosen
to give an ionic (kinetic) temperature $T_{\rm ion}=T_{\rm in}$.
While the fragmentation spectra may depend to some extent on the 
initialization procedure, the above procedure has been chosen 
to approximate the heavy-ion collision
experiments \cite{Grenoble95,Guet96,Guet97,Bergen98}.
The collision time is fast, of order 10~fs, so the ionic coordinates
$\{{\bf R}_{I}\}$ are effectively frozen during the ionization
process, as above.  The scaling factor $\lambda$ in the initial 
conditions is intended to approximate the extra energy 
``injected'' into the ionic system 
by relaxation of the valence electrons, which are excited during the 
collision.  Steps (iv) and (v) above effectively assume this relaxation
to be very rapid.  We discuss the relative sizes of relaxation 
and fragmentation times below.

\begin{figure}[tb]
\epsfxsize 6.5cm
\begin{center}
\epsfbox{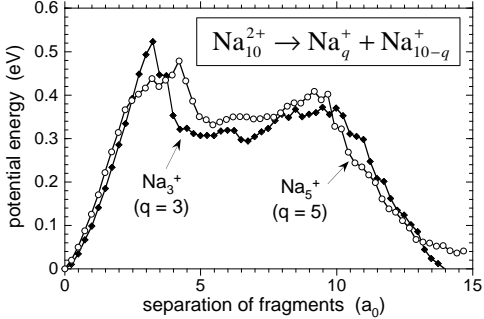}\end{center}
\caption{Fission barriers for two fission channels of Na$_{10}{}^{2+}$,
as a function of the separation of the centers of mass of the two
fragments, with zero corresponding to the initial cluster.}
\label{fig1}
\end{figure}

Although we are mainly concerned with barrierless fission
$X > 1$, to assess the accuracy of our KS approach, and to make
contact with previous experiment and theory, we first consider
briefly the binary fission of small doubly-charged clusters.  
Our lowest-energy
geometries for Na$_{N}$ ($4 \leq N \leq 8$) agree with previous
density-functional theory
(DFT) studies \cite{Martins83,Bar91a}, and our pseudopotential
is adjusted to give a good fit to ionization and atomization
energies for this size range, to within the scatter of previous
DFT results \cite{Martins83,Bar91a}.  We find doubly-charged clusters
Na$_{N}{}^{2+}$ to be unstable for $N \leq 6$, in agreement 
with Ref.\ \cite{Bar91a}.
In dynamical simulations with $M = 10$ initial 
conditions for each $N$ and with $T_{\rm in}=600$~K,  
Na$_{N}{}^{2+}$ ($5 \leq N \leq 10$) undergoes binary fission with 
Na$_{3}{}^{+}$ as the dominant fragment, in agreement
with experiment \cite{Brech91,TP92} and previous theory \cite{Bar91a}.  
Fission products other than Na$_{3}{}^{+}$ are found in only two cases: 
Na$_{6}{}^{2+} \rightarrow {\rm Na}_{5}{}^{+} + {\rm Na}{}^{+}$ 
with about a 20\% 
branching ratio, and Na$_{10}{}^{2+} \rightarrow 2{\rm Na}_{5}{}^{+}$ 
with about a 40\% branching ratio.  

To understand the result for
Na$_{10}{}^{2+}$ further, we show in Fig.\ \ref{fig1} the energy barriers
for the two observed fission channels, 
obtained by constrained energy minimization 
with the separation of the centers of mass of the two fragments
specified.  A double-humped fission barrier is found for each channel, 
and in the dynamics the cluster elongates to a ``precursor state''
where it may remain for several ps before fissioning, as also found
previously in Ref.\ \cite{Bar91a}.  Our barrier height for Na$_{3}{}^{+}$
emission is about 0.5~eV, in reasonable agreement with the 0.7~eV found
in Ref.\ \cite{Bar91a}.  The discrepancy may be due in part to
our phenomenological pseudopotential.

Let us now turn to a systematic study
of the effect of fissility on the fragmentation spectrum
for fissilities greater than one.  We shall consider the
fragmentation of Na$_{24}{}^{Q+}$
for $Q = 4$--8 ($X = 1.7$--6.7) with initial temperatures 
$T_{\rm in} = 400$~K, 800~K, and 1500~K.  
We find Na$_{24}{}^{Q+}$ to be unstable (at 0~K) 
in our KS model for $Q \geq 4$; 
Na$_{24}{}^{3+}$ ($X = 0.94$) is just stable, 
with a barrier of about 0.2~eV for removal of Na$_{3}{}^{+}$.
For each $Q$ and
$T_{\rm in}$, we run $M = 10$ simulations, each lasting up
to 5~ps ($Q=4$), 3~ps ($Q=5$--6), or 2~ps ($Q=7$--8).
A distinct final-state fragment Na$_{n}{}^{q+}$ is considered to have 
formed when all $n$ ions in it are separated from the remaining ions by 
more than a cutoff distance $r_{\rm cut} = 14.0$~a.u. Its charge $q$
is calculated as the total charge inside a box centered on the fragment
with a border of at least 7.0~a.u.\ from any ion.  Usually, $q$ is 
integral to better than a few percent, and the identification of the
fragment is unambiguous.  But this is not guaranteed by the KS
formalism: when two virtual orbitals centered on different clusters
are nearly degenerate and overlap, the resulting KS orbital may ``split'' 
between the two centers yielding fractional charges.  
This turns out to be particularly 
likely to happen with monomers or dimers emitted toward the end of a 
multifragmentation process.

In such cases, it is usually possible to assign integral charges 
unambiguously by interpreting the electron wavefunctions 
statistically.  A typical example would be
\begin{equation}
{\rm Na}_{24}{}^{8+} \rightarrow
4{\rm Na}^{+} + {\rm Na}_{18}{}^{2.85+} + {\rm Na}^{0.53+} + {\rm Na}^{0.70+}
. \label{eqn:fracq}
\end{equation}
The first four fragments emitted here are Na$^{+}$ with very close
to integer charge, but when the simulation is stopped (here after 2~ps),
the remaining fragments are fractionally charged.  We round the
large fragment up to Na$_{18}{}^{3+}$, and assume that the total remaining 
charge of 1.08 shared by the two monomers is to be interpreted, 
in a statistical 
sense, as ${\rm Na} + {\rm Na}^{+}$, with the probability
for finding the charge +1 on a particular monomer given
by the fractional charges.  In this way, we often find
neutral monomers or dimers emitted in the final stages of a 
multifragmentation (but never among the initial fragments),
which we interpret as evaporation from a hot residual fragment.  
Note that the slight excess of positive charge 
$\sum_{i} q_{i} = 8.08$ on the r.h.s.\ of Eq.\ (\ref{eqn:fracq}) is due to 
spillout of electron density from the boxes used to calculate
the total charge, which results in a slight underestimate of the 
negative electronic charge.
When simple rounding or charge redistribution among equivalent species 
does not give a clear assignment of integral charges, we discard the 
simulation, which was the case for less than 5\% of simulations.

\begin{figure}[tb]
\epsfxsize 8.0cm
\begin{center}\epsfbox{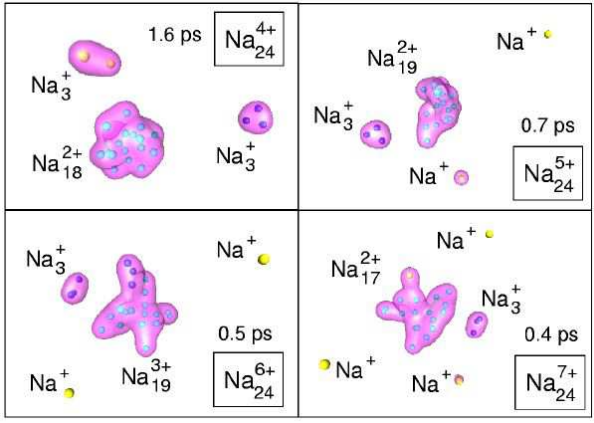}\end{center}
\caption{Snapshots of the Coulomb fission of Na$_{24}{}^{Q+}$ for $Q=4$--7
and an initial temperature $T_{\rm in}=800$~K.  The time $t$ for each 
snapshot (with $t=0$ the initial condition) is shown.}
\label{fig2}
\end{figure}

Some typical snapshots of Coulomb fission processes illustrating 
the main points are shown in Fig.\ \ref{fig2}.
For $Q=4$ ($X=1.7$) (and also for $Q=3$, $X=0.94$), we find mainly
sequential emission of Na$_{3}{}^{+}$ on a 1~ps time scale, with
only rarely Na$^{+}$ emission.  
Such emission continues until the large
residual fragment (which ultimately develops a fission barrier)
is too cool to emit further fragments,
at least on the time scale of our simulation, here 5~ps. 
(It is possible that with a longer simulation time we would 
occasionally observe emission of an additional singly-charged 
fragment.)
As $Q$ increases, we find emission of increasing 
quantities of Na$^{+}$, which is the dominant fragment for
$Q \geq 5$.  All singly charged fragments up to size six are observed
in some quantities.  

\begin{figure}[tb]
\epsfxsize  7.5cm
\begin{center}\epsfbox{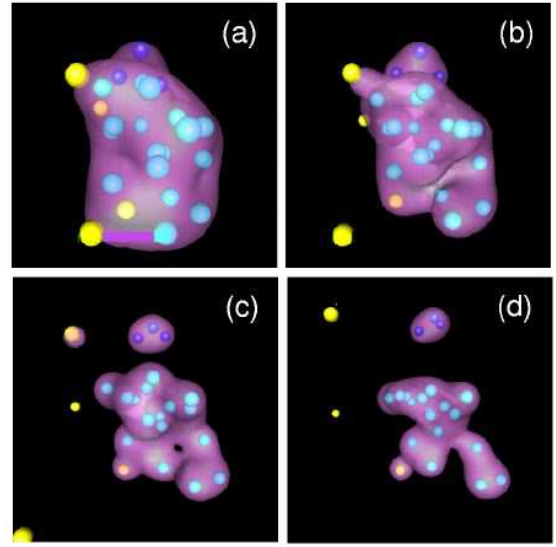}\end{center}
\caption{Coulomb fission of Na$_{24}{}^{7+}$ at times of (a) 0.1~ps, 
(b) 0.2~ps, (c) 0.3~ps, and (d) 0.45~ps.  Shown are isocontours
of the valence electron density, with Na$^{+}$ ions in the interior.}
\label{fig3}
\end{figure}

Figure \ref{fig3} shows in detail a 
disintegration of Na$_{24}{}^{7+}$ ($X=5.1$) as a function of time.  In 
Fig.\ \ref{fig3}(a), taken at $t=0.10$~ps, two Na$^{+}$ at the
top and bottom left are about to leave the cluster.  After
0.2~ps [Fig.\ \ref{fig3}(b)], a third Na$^{+}$ and a Na$_{3}{}^{+}$
start to leave from the rear.  At $t=0.3$~ps [Fig.\ \ref{fig3}(c)],
the residual cluster is already highly deformed, and eventually
emits a further Na$^{+}$ and Na$_{3}{}^{+}$ after about 0.7~ps, leaving
a Na$_{14}{}^{+}$ residue that remains stable up to 2~ps, when the
simulation was terminated.  The first few Na$^{+}$ ions emitted
at about $t \sim 0.1$~ps were initially at the surface, where the
valence electron density is low (the excess positive charge tends
to be located near the surface in a metallic cluster).  We conclude
that these ions were sufficiently weakly bound that
they they simply accelerated outwards starting at $t=0$.

\begin{figure}[b]
\epsfxsize 6.2cm
\begin{center}\epsfbox{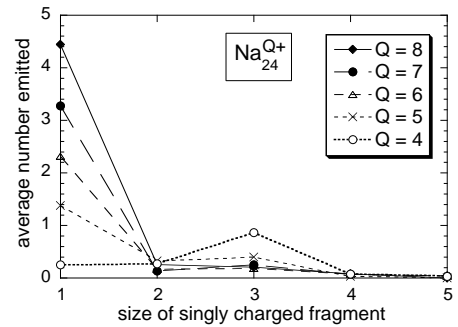}\end{center}
\caption{Average number of Na$_{n}{}^{+}$ fragments emitted per event
versus the size $n$ of the fragment, for the
Coulomb fission of Na$_{24}{}^{Q+}$ ($Q=4$--8).}
\label{fig4}
\end{figure}

The precise fragmentation pattern found in a given run for large $X$ 
is very sensitive to the initial condition, even for a given $T_{\rm in}$.  
However, averaging over initial conditions, we find the 
mean number of a given fragment emitted to be, within statistical error, 
essentially independent of $T_{\rm in}$ for $T_{\rm in}=400$~K,
800~K, and 1500~K.  This is perhaps not surprising, since the Coulomb energy
of Na$_{24}{}^{Q+}$ is $E_{\rm Coul}
\approx 1.2 Q^{2}$~eV, while the ionic kinetic energy is only $E_{\rm kin}
\approx 0.3$~eV per 100~K and is thus small compared to $E_{\rm Coul}$
for all cases considered here.  Therefore, in Fig.\ \ref{fig4} we 
have combined our results for the three $T_{\rm in}$, and show
the average number of singly-charged fragments Na$_{n}{}^{+}$ 
($n=1$--5) emitted per fragmentation, averaged now over
30 initial conditions for each parent charge $Q$.
The most striking trend observed in Fig.\ \ref{fig4} is the 
smooth increase in number of Na$^{+}$ emitted with increasing $Q$,
accompanied by a reduction in the number of Na$_{3}{}^{+}$ emitted.  
For Na$_{24}{}^{4+}$ ($X = 1.7$), Na$_{3}{}^{+}$ dominates; 
for Na$_{24}{}^{8+}$ ($X = 6.7$), on average 20 times more Na$^{+}$
are emitted than Na$_{3}{}^{+}$.  

The average time elapsed before the emission of the first fragment varies
from 0.1--0.2~ps for $Q=6$--8, where the first fragment is nearly
always Na$^{+}$, to 0.4--0.6~ps for $Q=4$--5, where Na$_{3}{}^{+}$
is most often the first fragment.  A combined electronic-ionic 
dynamical study within time-dependent KS \cite{Calvay98} 
suggests that electron-ion relaxation times $\tau_r$ may be 
of order $\tau_r \leq 100$~fs,
and thus competitive with the time scale $\tau_f$ of some faster 
fragmentation processes ($\tau_f \geq 50$~fs).  However, as we have
seen, these faster processes involve immediate acceleration of Na$^{+}$ 
away from the surface region of the cluster, a process which does not
require electron-ion relaxation to occur.  Moreover, our
mean fragment spectra are found to be essentially independent of 
$T_{\rm in}$ up to at least $T_{\rm in}=1500$~K.  We do not believe, 
therefore, 
that our approximate treatment of electron-ion relaxation via 
the parameters $\lambda$ and $T_{\rm in}$ will lead to a significant 
qualitative error in our results.

The trend shown in Fig.\ \ref{fig4} is consistent 
with the observation in a heavy-ion collision experiment \cite{Guet97} that 
Na$^{+}$ dominates the inclusive small 
fragment spectrum when Xe$^{20+}$ is the projectile, while Na$_{3}{}^{+}$
dominates when Ar$^{3+}$ is the projectile, since the heavier ion Xe$^{20+}$
should produce clusters Na$_{N}{}^{Q+}$ with higher charge $Q$.  Our
results are consistent also with another collision experiment 
\cite{Bergen98}, in which Na$_{3}{}^{+}$ is found to dominate 
in events with a single light fragment, while Na$^{+}$ dominates in events 
with a multiplicity $p$ of light fragments $p \geq 2$.  We conclude
that in the latter events one is observing multifragmentation of parent 
clusters with $X > 1$.

We note that the preference for Na$^{+}$ emission 
for $X \gg 1$ is qualitatively consistent with simple considerations of the 
total energy released ($Q$-value) in a charged liquid-drop model \cite{Sven97},
which favors the distribution of the parent charge over many small fragments.  
Also, in a statistical model \cite{GrossIII} that assumes an ergodic
distribution of fission fragments for a system confined to a small volume, 
evidence was found for a first-order ``fragmentation phase transition'' 
from a regime with a large residual fragment at low excitation energies to
a regime with only small fragments at high excitation energies.  We do 
not observe clear evidence for such an effect in the present data, 
although our statistics
and range of study are too limited.  However, we note that, in
addition to providing detailed spatial and temporal information, the present 
dynamical approach avoids the ergodic assumption in the statistical model.

Our test system Na$_{24}{}^{Q+}$ is perhaps too small to be able to observe jets 
unambiguously.  We observe only small fragments, which tend to be emitted 
isotropically, although the cluster has a tendency to elongate during 
the process (see Figs.\ \ref{fig2} and \ref{fig3}).  We stress that these
conclusions pertain to the regime of barrierless fission $X > 1$.  We
have also searched for symmetric fission when $X < 1$, where 
experimental evidence exists for the emission of large fission
fragments \cite{Brech96entrop}.  We found symmetric fission for
Na$_{10}{}^{2+}$, and as a rare event ($<10$\%) for Na$_{18}{}^{2+}$,
which has a favorable closed-shell final state Na$_{9}{}^{+}$.  
However, in a low-statistics study of Na$_{40}{}^{4+}$ ($X = 1$), we observed 
only sequential Na$_{3}{}^{+}$ emission (and for Na$_{40}{}^{8+}$ we
observed predominantly Na$^{+}$ emission).

RZ would like to acknowledge the support of the Indo-French Center for the
Promotion of Advanced Research under project 1901-1.

\end{document}